# Molecular Inverse-Design Platform for Material Industries

Seiji Takeda[1*], Toshiyuki Hama[1], Hsiang-Han Hsu[1],
Victoria A. Piunova[2], Dmitry Zubarev[2], Daniel P. Sanders[2], Jed W. Pitera[2],
Makoto Kogoh[3], Takumi Hongo[3], Yenwei Cheng[3], Wolf Bocanett[3], Hideaki Nakashika[3],
Akihiro Fujita[4], Yuta Tsuchiya[4], Katsuhiko Hino[4], Kentaro Yano[4],
Shuichi Hirose[5], Hiroki Toda[5], Yasumitsu Orii[5], Daiju Nakano[1]

[1] IBM Research – Tokyo, [2] IBM Almaden Research Center, [3] IBM Garage, Tokyo Laboratory,
[4] HAYASHIBARA Co., Ltd., [5] NAGASE Co., Ltd.

{seijitkd, hama, hhhsu, dnakano, makotox, nakashika}@jp.ibm.com,
{vapiunov, dsand, pitera}@us.ibm.com, {dmitry.zubarev, takumi.hongo, yenwei.cheng, wolf.bocanett}@ibm.com,
{akihiro.fujita, yuta.tsuchiya, katsuhiko.hino, kentaro.yano}@hb.nagase.co.jp,
{hiroki.toda, shuichi.hirose, yasumitsu.orii}@nagase.co.jp

## ABSTRACT

The discovery of new materials has been the essential force which brings a discontinuous improvement to industrial products' performance. However, the extra-vast combinatorial design space of material structures exceeds human experts' capability to explore all, thereby hampering material development. In this paper, we present a material industry-oriented web platform of an AI-driven molecular inverse-design system, which automatically designs brand new molecular structures rapidly and diversely. Different from existing inverse-design solutions, in this system, the combination of substructure-based feature encoding and molecular graph generation algorithms allows a user to gain high-speed, interpretable, and customizable design process. Also, a hierarchical data structure and user-oriented UI provide a flexible and intuitive workflow. The system is deployed on IBM's and our client's cloud servers and has been used by 5 partner companies. To illustrate actual industrial use cases, we exhibit inverse-design of sugar and dye molecules, that were carried out by experimental chemists in those client companies. Compared to a general human chemist's standard performance, the molecular design speed was accelerated more than 10 times, and greatly increased variety was observed in the inverse-designed molecules without loss of chemical realism.

* Corresponding Author

## CCS CONCEPT

• **Applied computing** →Physical science and engineering → *Chemistry*; • **Computing methodologies** → Machine learning → Machine learning algorithms →*Feature selection*; • **Mathematics of computing** → Discrete mathematics → Graph theory → *Graph enumeration*;

## 1. INTRODUCTION

The discovery of new materials has been the key driving force behind discontinuous development across many industrial domains including those for automobile, aircraft, pharmaceutical, etc. In particular, a great deal of attention has been centered on organic molecular design owing to its broad range of applications and highly complex structural variety. In industrial product development, the variety of available materials species constrains the boundary of product design, and therefore determines the limitations of product's performance. Designing "tailored" materials that possess desired characteristics is a long-awaited capability that can unleash a product's performance from the envelope defined by off the shelf available materials.

Materials development is the process of solving a complex multi-objective combinatorial problem in terms of material structures. Today's development is proceeded by running numerous trial-and-error cycles consisting of material structure design, screening by simulation, and experimental validation, mostly on the basis by knowledge, experience, and intuition of human experts. However, even when limited to small molecules, the estimated number of molecular structures' combinatorial patterns significantly exceeds $10^{60}$ [1]. In this extra-vast parameter space, experts in each material domain can only explore around a tiny space relying on trial-and-error strategy. Therefore, the material development period typically takes as long as 10-20 years, and most of material innovation falls into minor incremental improvement of today's lead candidate. On that background, many R&D activities have been carried out to leverage data science, machine learning, and AI so that design speed is accelerated, and diversity of designed material structures can be expanded beyond each individual expert's knowledge and potential biases.

## 2. RELATED WORK AND OVERVIEW

In this section, we will introduce the state-of-the-art of today's AI-driven material design solutions especially focusing on molecular design, define the issues to solve, and describe our new contribution.

### 2.1. State-of-the-Art

The first step in AI-driven material discovery is to build an accurate regression model to predict molecule's properties. For that purpose, historically a variety of feature vectors to encode a molecule have been reported. To capture an organic molecule's graph topology, the most popular is the fingerprint [2]. A recent more powerful method is Graph Convolutional Neural Network (GCNN) [3-5]. Meanwhile, 3D molecular structure is captured by methods like the Coulomb Matrix [6]. Besides such explicit feature engineering, recently representation learning to handle "raw" structural information has been reported. In [7-9], SMILES (Simplified Molecular Line Entry System) string, that is a direct string representation of a molecular graph topology, is embedded in a feature space by Deep Neural Network (DNN). In [10], a 2D image of a molecular graph is directly input to a DNN. The feature spaces built by these schemes are used in structure design processes by two main approaches: Virtual Screening or Inverse-Design.

Virtual Screening, which is commonly used in the pharmaceutical sector, is a popular approach in material design. In this approach, firstly a set of diverse molecular structures with unknown property values are exhaustively generated by genetic algorithms [11], fragment-based generation [12], SMILES generation [13] etc. Those structures are filtered by a property prediction model so that only a set of promising structures are screened as candidate structures [14]. This approach is useful given that users can spare substantial computational resources required for training models and generating a large set of structures. However, this approach is somewhat limiting due to its exhaustive search concept.

A recent promising approach is Inverse-Design; generating candidate molecular structures by inverse-solving the property prediction model starting with user-demanded target property values. In [7-9], combinations of Variational Auto Encoder (VAE), Recurrent Neural Network (RNN), Reinforcement Learning (RL), and Bayesian optimization have been realized to generate new molecular structures in SMILES strings satisfying the targeted properties. In [15], SMILES were generated by combining RNN with Monte Carlo Tree Search (MCTS). Besides SMILES generation, in [16], a molecular graph matrix is generated by RL and Generative Adversarial Network (GAN). The state-of-the-art of inverse materials design using DNN is summarized in [17].

### 2.2 Issues of Existing Solutions

Some of the above methods are packaged and available on GitHub. Those libraries are outstanding landmarks in the history of AI-driven material discovery, however, they are not fully architected for the real material development in industry for the following 4 reasons. (1) To make those DNN generative models to generate structures *correctly* on the aspects of SMILES grammar's validity, structural "similarity", chemical feasibility etc, heavy load training of the generator by using a large data set is inevitable. However, in real material applications, the available data in each advanced domain are extremely small; on the order of 10 to 100. (2) Most DNN models are black-box; feature vectors in the latent space is uninterpretable, therefore tuning of the model is difficult for chemists. (3) The tool architecture is often not designed to support AI version of trial-and-error; tune hyper parameters, build models, review results, give feedback, ingest hand-crafted rules, store results and share with collaborators, etc. (4) It is a big barrier for experimental chemists to carry out those tasks on a Command Line Interface in their daily work. For the above reasons, the state-of-the-art does not yet provide a truly material industry-oriented inverse-design tool or service.

### 2.3 Contributions

Given the above, our contributions include the following.
- We developed a molecular inverse-design system, where chemists can design brand new chemical structures satisfying targeted property values.
- We developed feature encoding scheme and graph generation algorithms by a substructure-based approach, that is not data-hungry. They provide interpretable information so that a user can customize features and generation processes by introducing domain knowledge.
- We developed a system architecture in which process histories are stored in a hierarchical structure so that a user can perform trial-and-error flexibly.
- We developed a Graphical User Interface for the system as a web application, that allows intuitive utility to a user.
- We exhibit two industrial use cases of sugar and dye design, carried out by our platform users. It was confirmed that our system outperformed human chemists in speed and diversity of molecular design.

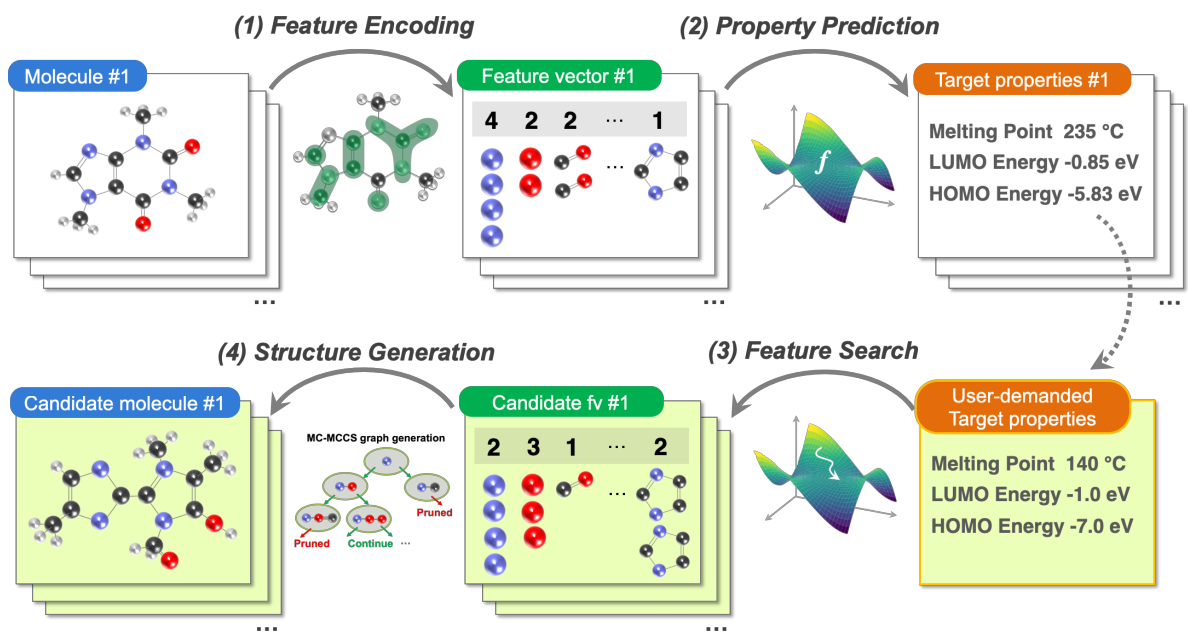

Figure 1: Fundamental workflow of Molecular Inverse-Design system.

## 3. METHODS

Conducting interviews with more than 50 materials and chemical companies, we architected an algorithm workflow aiming at highly practical use in chemists' daily material design work. Critical requirements are, first, for the system to mine insights from industrially realistic small datasets. Second, high interpretability and tunability in both feature encoding and structure generation processes is needed. Third, high domain-adaptability should be achieved by allowing a user to flexibly customize the system to include domain knowledge. The fundamental workflow consists of (0) data input, (1) feature encoding, (2) prediction model building, (3) feature search, and (4) structure generation, as described in Figure 1. Detailed information for each step is presented below.

### 3.1 Feature Encoding

For practical use especially in chemical industries, a feature vector should be interpretable, customizable, invertible so that molecular structures can be generated from it, and of course sufficient to model molecular structures accurately. To satisfy those requirements, we developed a substructure-based encoding scheme, that is highly compatible with graph generation algorithms.

A molecule is a graph structure consisting of nodes and edges that correspond to atoms (e.g. carbon) and chemical bonds (e.g. double bond), respectively. A molecular graph is comprised of a variety of subgraphs, namely substructures. The concept of our encoding scheme is to vectorize frequencies of substructures appearing in a molecule in the similar manner as bag-of-words in natural language processing. First, substructures appearing in an entire dataset are extracted and then compiled as a list. Next, the numbers of the listed substructures appearing in each molecule are counted, and those numbers are arrayed as a feature set. Different levels of substructures (size, abstractness, extraction scheme,) yield corresponding feature sets, from which a user selects appropriate ones, and then concatenate them to a feature vector representing the molecule.

The first advantage of this method is that the configured feature space is fully interpretable for chemists. They can review features' contents and customize them by adding/deleting features. Also, a user can understand which part of molecular structure has impact on chemical properties, important insights for material scientists to consider design direction. The second advantage is that this feature representation can, when combined with a graph generation algorithm, be inverted to concrete molecular structures, different from other rule-based features [2-6]. Also, the inversion process is not stochastic as VAE or GAN [7-9], but fully controllable; from sparse to exhaustive level of generation.

### 3.2 Property Prediction

A regression model to predict target properties is built by using the above feature vector. The best model and hyper parameter set are automatically selected by default, but a user can manually select a model from Lasso, Ridge, ElasticNet, Kernel Ridge, Random Forest, and Support Vector regression models with hyper parameters. Built models are stored in the system so that a user can select one for the next Feature Search step.

Different from other machine learning tasks, DNN is not always the best option, because in material industries, the available data labeled by meaningful target properties are much smaller than cases of image recognition, speech recognition, text mining and so on. Some exceptional cases are where target properties are very low-level ones (e.g. energy information), so properties of a myriad number of molecules can be calculated by physical simulation. In some recent papers, transfer learning is leveraged to complement this data shortage [5], but pursuing this direction is a future challenge.

### 3.3 Feature Search

A user sets targeted property values and runs the Feature Search process to search out candidate feature vectors satisfying the targets on the regression model. For the search process, we developed Molecular Constrained Particle Swarm Optimization (MC-PSO) algorithm, that constrains search space by both rule-based and data-driven bounds. PSO is a powerful optimization algorithm by itself, but constraints on the feature space considering each vector's molecular feasibility (meaning, whether a feature vector can be inverted to a concrete molecular structure) are crucially important, because feasible solutions distribute in substantially narrow manifolds.

Some chemical constraints can be explicitly given by rules. For example, in a molecule, the number of oxygen atoms (O) should be apparently larger than or equal to the number of substructures hydroxyl group (-OH). Likewise, several explicit rules are set. Meanwhile, other constraints are implicitly determined by unknown relationships between the numbers of substructures, therefore we exploited a data-driven approach. We preliminarily generate small graphs exhaustively, and map them to a 4-dimensional space, whose each dimension represents the number of subgraphs having 1, 2, 3, and 4 edges respectively. The mapped data points form discrete manifolds, that are molecular "feasible" area, in which molecular structures can be generated. Running PSO, candidate feature vectors are mapped to this 4-D space, where only the ones dropped in those area are screened as candidates. This chemical constraint approach increased the "success rate" to search out feasible feature vectors by more than 10-fold. By customizing the search range of each feature, a user can inject domain knowledge. For example, if aromatic rings are not desired, search ranges of all the substructures including aromatic rings can be set [0, 0].

### 3.4 Structure Generation

As the last step, the sampled candidate feature vectors are expanded to concrete molecular structures by a high-speed graph generation algorithm. The algorithm repeats cycles of connecting fragments (i.e. atoms, rings, and substructures), and screening by feature element's value. The most important is to generate structures both exhaustively (without omission) and efficiently (without isomorphic duplication). Connecting fragments easily brings on exponential divergence in combination patterns, in which most of structures are isomorphic duplications. Therefore, it is important to prune

---

**Algorithm 1:** Main algorithm of structure generation
**Input:** #Atoms // dictionary of atoms and numbers
#Fragments // dictionary of fragments and ranges
**Output:** molecular structures as graphs
1: **generate**(#Atoms, #Fragments)
2:   **Foreach** a **in** #Atoms.keys
3:     g = <{a}, {}> /* new graph */
4:     /* get labeling and automorphism */
5:     label, auto_m = *canonical_labeling*(g)
6:     #Atoms0 = *decrement* #Atoms[a]
7:     *augment*(g, auto_m, #Atoms0, #Fragments)
8:   **End**

**Algorithm 2 :** Building a molecular graph
1: **augment**(g, auto_m, #Atoms, #Fragments)
2:   **If** *satisfy_constraint*(g, #Fragments)
3:     /* solution found */
3:     *output* g
4:   **Endif**
5:   **If** *check_termination*(g, #Aoms, #Fragments)
6:     /* no possibility in future */
7:     **return**
8:   **Endif**
9:   /* augment a graph by adding a new atom */
10:  **Foreach** orbit **in** auto_m
11:    v = *representative of* orbit
12:    **Foreach** a **in** #Atoms.keys
13:      **If** #Atoms[a] > 0
14:        **Foreach** bond **in** {1, 2, 3}
15:          **If** bond + *degree*(v) <= *valence*(v)
16:            e = *edge*(v, a, bond)
17:            g0 = <g.V+{a}, g.E+{e}>
18:            /* get labeling and automorphism */
19:            label, am = *canonical_labeling*(g0)
20:            /* check canonical augmentation */
21:            **If** label[a] == 0
22:              #Atoms0 = *decrement* #Atoms[a]
23:              *augment*(g0, am, #Atoms0, #Fragments)
24:            **Endif**
25:          **Endif**
26:        **End**
27:      **Endif**
28:    **End**
29:  **End**

**Algorithm 3 :** Graph check by fragment's number
1: **satisfy_constraint**(g, #Fragments)
2:   /* check acceptable range of fragments */
3:   **Foreach** fragment **in** #Fragments.keys
4:     range = #Fragments[fragment]
5:     count = *count_fragment*(g, fragment)
6:     **If** not range.*include*(count)
7:       **return** False
8:     **Endif**
9:   **End**
10:  **return** True

generation paths before generating duplicated structures. For that purpose, we developed Molecular Customized McKay's Canonical Construction Path (MC-MCCP) algorithm. The original MCCP generates graphs by connecting vetices, in which the "canonical label" of a graph in each generation steps is calculated to judge whether the graph can be further grown or not from the viewpoint of isomorphism [18-20], as implemented in MOLGEN; a web application to generate molecules by simple conditions on the number of atoms and bonds [19]. In our case, the problem is more complex; molecular graphs include a variety of rings and substructure patterns, therefore generation paths should be pruned by forecasting graph feasibility, not considering only isomorphism.

The algorithms of MC-MCCP are summarized in Algorithm 1-3, where detail of the function *canonical_labeling* is described in [20]. Major differences from the original MCCP are, (1) each of rings and substructures are dealt with as a single vertex in *#Fragments*, making it possible to generate larger graphs, (2) structural feasibility along the current generation path is forecasted by *check_termination*, where remaining fragment's and feature vector's values are compared, and non-promising paths are pruned early, and (3) the numbers of fragments are checked at each step in *satisfy_constraint.* Those improvements made MCCP applicable to inverse-design of large and complex molecular structures. The system accepts hand-crafted rules of structural constraints; range of the number of substructures, user-defined fragment patterns, so that chemical knowledge can be included.

## 3.5 Customization for Domains

Presented above are our basic and general purpose molecular inverse-design algorithm. This is useful by itself as described in section 5, however, in actual industries, each material domain imposes complex structural rules on molecules. For example, in many domains only molecular components attached to a common template should be designed.  In order to customize the system to those structural rules, we implemented a variety of user-definable customization methods. They are mainly classified to three categories; (1) template customizer : a user can define any structural templates to extract molecular components to design, (2) feature customizer : a user can concatenate any additional data as features (e.g. composition ratio, etc), (3) efficient inverse-design scheme for several components in a molecule. Each of those categories consists of several detailed methods. By combining them, a user can customize the system to broad organic material domains.

## 4. DEPLOYMENT

The above algorithms are implemented as a comprehensive material design system, and deployed on IBM's and clients' clouds as-a-service. The system is architected to provide high flexibility in configuring a workflow and customizing functions

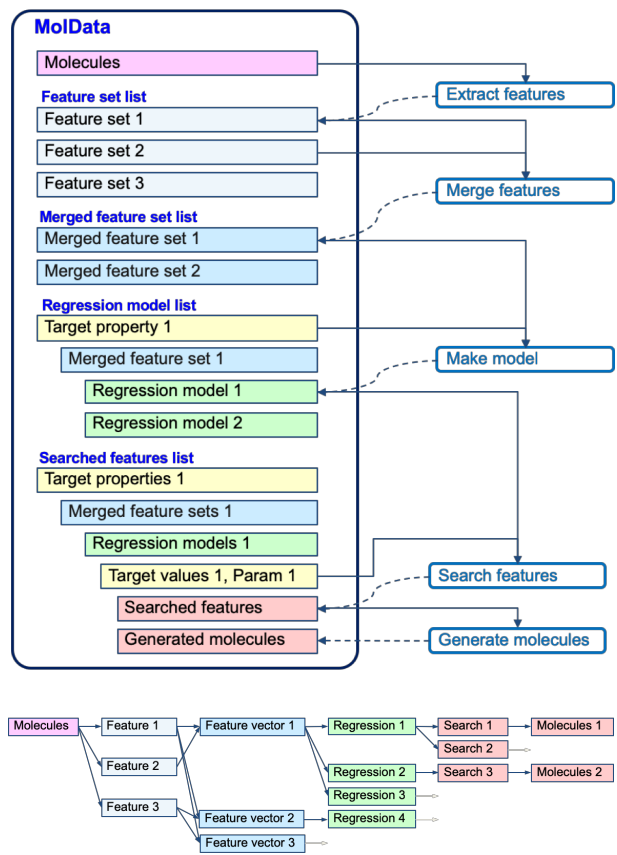

**Figure 2: The hierarchical architecture of MolData object (Top) and a flexible workflow (Bottom).**

for specific material domains. In this section we will introduce about implementation detail.

### 4.1 Data Architecture

In the actual industrial use, the workflow a user experiences in inverse-design (and any other AI-driven services) is not straightforward. Not only running auto-optimization, a user needs to try different data subsets, features, models, hyper parameters, etc. Our system is designed to allow a user to configure such flexible trial-and-error processes.

Input data is uploaded in the CSV format, which consists of pairs of molecular structures (SMILES) and associated properties.  The uploaded data is converted to "MolData", which is an object having a hierarchical structure, which stores input data and all the interim results as shown in Figure 2. On the MolData object, each time a user runs a method, the output result is stored as an object which can be selected as input for the next method. For example, by running "Extract features" method with different parameters, corresponding outputs "Feature set 1, 2, 3.." are stored. The user can select some of them to merge into a "Merged feature set 1", that can be used as feature vectors for building regression models with different parameters. In this manner, result objects are stored hierarchically with deep branches. So a user can look over the design status, and switch to any result object from which design process can be continued as described in Figure 2.

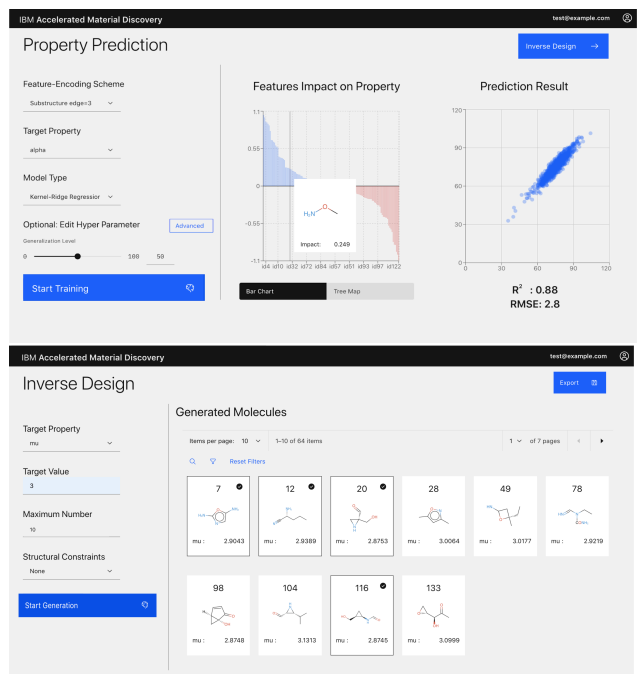

**Figure 3: UI screenshots of the web application version of Molecular Inverse-Design system.**

## 4.2 Service Platform

The system was implemented as a Python package. Some of the low-level functions were developed with open source packages: RDKit for drawing chemical structures from SMILES and Scikit-learn for building common regression models. The service has two interfaces according to users' IT skill levels; one is a Python script-based interface on Jupyter Hub, and the other is a GUI-based web application. On the script-based version, a user can benefit from more than 100 detailed methods. On the web application, a clean UI screen provides intuitive usability to non-IT-skilled users. A user is visually guided to a workflow, so they can carry out inverse design by only a few mouse-clicks. Most of the method details are suppressed in default, but in "advanced" mode, they are gradually exposed. Some of the UI screenshots are shown in Figure 3.

Today, the system is deployed on IBM's cloud and is in daily use by material scientists from our partner companies. Also, a custom-made version is deployed on the cloud for material discovery service operated by NAGASE Co., Ltd., where many chemists are using our to system to design sugar and color materials. Technical details of those designs will be described in Chapter 6.

## 5. BASIC PERFORMANCE

In this section, we demonstrate the inverse-design of new molecular structures by using a public dataset. We will exhibit the performance of our system in the basic use case scenario.

### 5.1 Data

We used a dataset extracted from the QM9 dataset, which is a small organic molecules dataset consisting of 134,000 (134k) molecular structures and their chemical properties calculated by the Density Functional Theory (DFT) method [21]. Considering the real industrial situation where the amount of data is mostly small, we extracted a subset of 1,000 molecule (1k) from QM9. Using this dataset, we perform inverse-design focusing on two target properties; LUMO energy ($E_{LUMO}$) and Energy gap ($E_{gap}$), that are important electronic properties that govern performance of organic semiconductor devices. Distribution of those properties of the 1k molecules are shown with molecular structures in Figure 4.

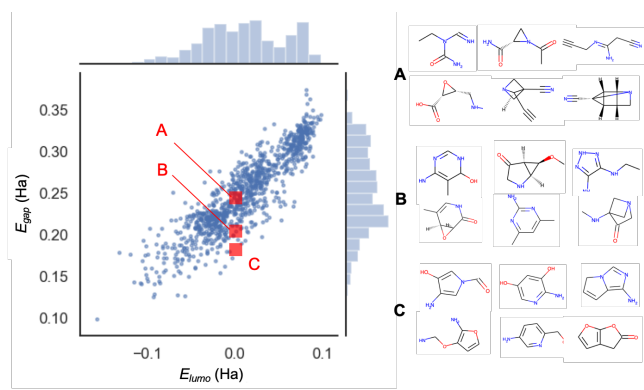

**Figure 4: Distribution of properties of the original data set. Red squares represent the target property areas. Randomly selected molecular structures included in the vicinity of those areas; A, B, and C are also shown.**

### 5.2 Experimental Results

First, the 1k molecular structures were encoded to 136-dimensional feature vectors, where each element represents the number of heavy atoms, rings, aromatic rings, and substructures. Next, the system builds regression models to predict $E_{lumo}$ and $E_{gap}$ respectively. For simplification, we carried out three regression methods; Lasso, Ridge, and ElasticNet with changing hyperparameters. To select the best model, we evaluated each model by average $R^2$ (decision coefficient) of test sets split out in 10-fold cross validation. For each model, we swept $L_1$ and $L_2$ penalty terms from $10^{-4}$ to $10^2$ with logarithmic increment. Considering the small number of samples, reasonable accuracies were confirmed as summarized in Table 1. As the best model, we selected Lasso regression model. Dimensions of feature vectors were slightly reduced to

**Table 1: Regression scores to predict the target property values ($E_{lumo}$, $E_{gap}$).**

|  | $E_{lumo}$ | | $E_{gap}$ | |
|---|---|---|---|---|
|  | $R^2$ | RMSE | $R^2$ | RMSE |
| Lasso | 0.842 | 0.017 | 0.793 | 0.018 |
| Ridge | 0.825 | 0.018 | 0.779 | 0.021 |
| ElasticNet | 0.839 | 0.017 | 0.783 | 0.020 |

**Table 2: Generation results about the number of candidate feature vectors, molecules, and required time.**

| Target ($E_{lumo}$, $E_{gap}$) | Feature Vectors | Generated molecules | Elapsed time (min) | Time/molecule (min) |
|---|---|---|---|---|
| (0.0, 0.25) | 8 | 103 | 95 | 0.92 |
| (0.0, 0.20) | 4 | 43 | 161 | 3.74 |
| (0.0, 0.175) | 7 | 90 | 615 | 6.83 |

105 and 112 for $E_{lumo}$ and $E_{gap}$ by feature selection. By making a union of both features, they were merged to a 117-dimensional feature vector set for running the MC-PSO.

Setting targets of ($E_{lumo}$, $E_{gap}$) to (0.0, 0.175), (0.0, 0.20), and (0.0 ,0.25) Hartree energy (Ha), we ran Feature Search. For each target, the search range with +/-σ of the predicted divergence was automatically set. Running MC-PSO, we obtained several number of feature vector candidates for each target. At the last step, we ran Structure Generation process for those feature vectors. For each searched out feature vector, structure generation was terminated after 20 molecules were generated. The number of searched out feasible feature vectors from which structures were successfully generated, and molecular structures obtained are shown with elapsed time in Table 2. Comparing with the standard chemist's design speed, which ranges in 0.1–3.0 day/molecule, an extreme acceleration is confirmed; more than 100 times. It is observed that generation speeds decrease when the original data points are sparse in the targeted area, but still the acceleration exceeds 10 times. Some of the molecular structures inverse-designed for the targets are shown in Figure 5. Comparing with Figure 4, structural similarities are confirmed in each target. As reported in [15], other DNN-based generative models' generation speeds range in 0.01-10 min/molecule. Our system's generation speed itself looks not very competitive with those rates, however, owing to the algorithmic nature, our system eliminates significant overheads; several hours to even 1 day required to train an encoder and a decoder (generator) with huge amount of dataset. This advantage significantly reduces the total lead time in inverse-design.

Finally, in order to validate the accuracy of our system, we carried out DFT simulation to compute the actual $E_{lumo}$ and $E_{gap}$ of the generated structures. Distributions of the calculated values for each target are exhibited in Figure 6. The deviations in simulated energies of designed structures reflect the model's accuracy; RMSE ~0.018 (Ha), and DFT simulation's accuracy. Except inevitable deviation originating from the validation itself using DFT simulation, those results proved the capability of our system to inverse-design organic molecules having targeted properties. In this section, the problem setting using QM9 is artificial to evaluate the basic performance. In the next section, we will exhibit the practical use case in material companies to solve real industrial problems.

($E_{lumo}$, $E_{gap}$) = (0.0, 0.25)

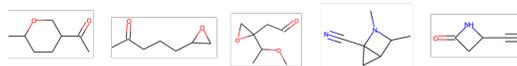

($E_{lumo}$, $E_{gap}$) = (0.0, 0.20)

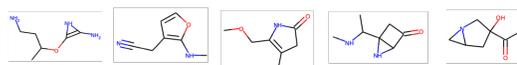

($E_{lumo}$, $E_{gap}$) = (0.0, 0.175)

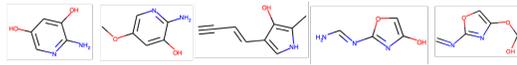

**Figure 5: A part of molecular structures inverse-designed for the three ($E_{lumo}$, $E_{gap}$) targets.**

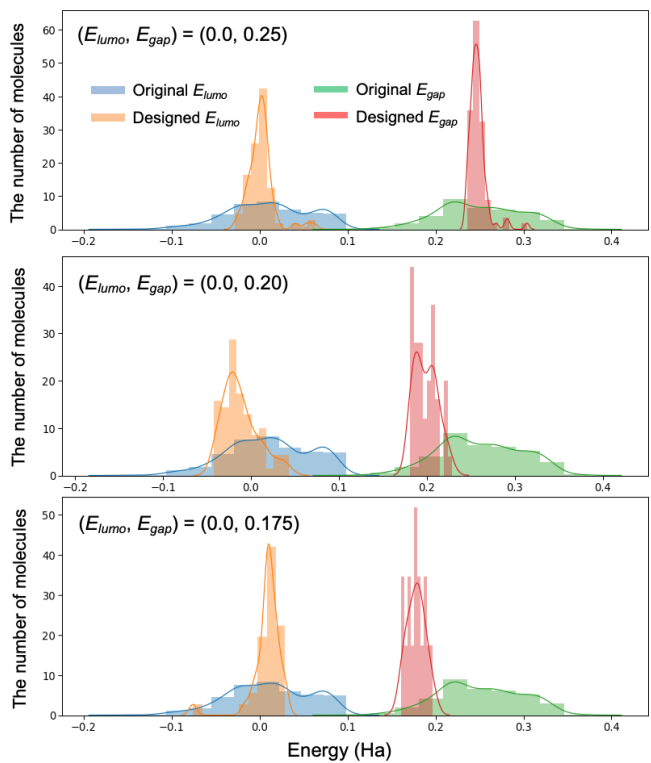

**Figure 6: Distributions of ($E_{lumo}$, $E_{gap}$) associated with the original dataset and DFT simulation results of the inverse-designed molecular structures.**

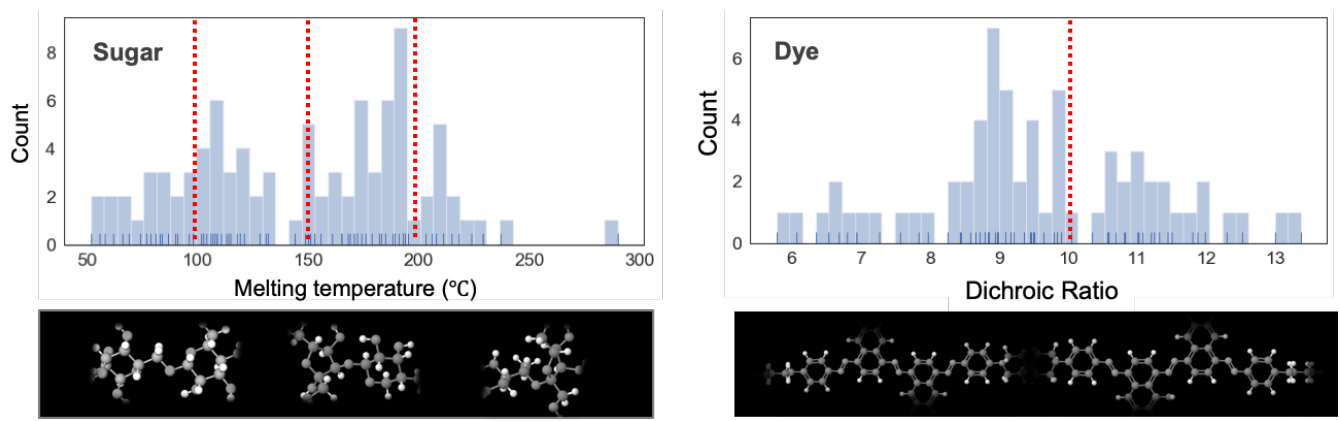

**Figure 7: The histograms represent the distribution of properties of the original data set. Red lines represent the target property values. The bottom images are partial structures of inverse-designed molecules.**

## 6. INDUSTRIAL USE CASE

Our system is implemented on the cloud server of NAGASE Co. Ltd., on which NAGASE's material discovery service for its client and subsidiary companies including HAYASHIBARA Co., Ltd. is running. The system deployed in this service was highly customized for specific product domains; sugar (saccharide) and dye. In addition to the general customization described in 3.5, we implemented several domain-specific new methods; stereochemical feature encoding, repeating units design, etc. Today, the system is in service to more than 50 experimental chemists. In this section, we will present those practical use case results limited to only disclosable public content.

### 6.1 Sugar Design

In the food industry, discovery of new ingredients adding desirable textures and flavors to formulated foods is a long-awaited solution. In particular, crucially important is to design sugar structures having properties in terms of melting temperature, sweetness etc. However, the number of configurable sugar structures exceeds over $10^9$ which is impossible to for only human chemists to explore. Thus, to accelerate saccharide discovery, experimental chemists of HAYASHIBARA collected their own proprietary experimental data to train our system. The data set includes 91 samples about saccharide structures and melting temperatures. Distribution of the melting temperatures is shown in Figure 7 with three targeted values.

Due to HAYASHIBARA's confidentiality, only a part of generated structures is presented in Figure 7, but two significant impacts were confirmed. First, despite that the original data didn't include any glycoside types (meaning, saccharide attached by other organic compounds), generated structures include many glycoside types, therefore the structural variation was more enriched comparing with the original dataset. This result proves the capability of our system to design beyond the scope envisioned by chemists. The second impact is that with increase of targeted melting temperature, inverse-designed molecules were confirmed to include more disaccharide and trisaccharide structures. This tendency is consistent with the common hypothesis in chemistry; sugar's melting temperature is affected by hydrogen bonding forces between hydroxyl groups, that increase with molecule's size and complexity. This fact highlights the chemical rationality of our system's design direction.

### 6.2 Dye Design

Azo dye is a well-known material, which changes its absorbance profile in accordance with spatial direction in a medium. Thus, the transparency of a mixture containing azo dyes and liquid crystals could be modulated by applying electric voltage. Azo dyes are widely used in liquid crystal panels, smart windows, etc. The most important physical property of an azo dye, which defines its application, is Dichroic Ratio (DR). However, complex atomistic interactions in a molecule and non-standardized data across institutes make the design process difficult. On that basis, 250 proprietary experimental data samples about dye structure and DR were collected and ingested to our system. HAYASHIBARA's chemists targeted DR=10 in the wide range of DR distribution as shown in Figure 7, then performed inverse-design.

More than 20 structures were generated. Part of the designed structures are shown in Figure 7. Similar to sugar's case, the variety in structural size and complexity was increased relative to the original dataset.

### 6.3 Lessons Learned from The Industrial Deployment

On the both sugar and dye cases, 10 to 100 molecular structures were generated by inverse-design carried out in only 1-day-long operation including all the end-to-end processes from data input to structure generation. This is a significant improvement in lead time to design the first pass candidate structures for validation. Standard chemist's lead times, estimated by HAYASHIBARA, are 3days/1molecule and 4days/1molecule for sugar and dye respectively. Also, it was confirmed that the enriched structural diversity in the inverse-design results brings about out-of-the-box structures, expanding the boundaries created by unconscious biases in scientists and datasets. Those structures were confirmed their chemical synthesizability, and further structural improvement and discovery are ongoing to reach the product level.

In order to accelerate the development of AI-driven materials discovery for real industries, we share some lessons learned from this deployment case. First, the system should not always count on "big data". Materials companies work on the state-of-the-art advanced materials, where data are natively rare and unavailable in public documents. One possibility to leverage big data is adopting transfer learning to train the fundamental model by public big data, and fine-tune with small proprietary data. Second, we conformed that highly domain-customizable AI solutions are required for material discovery. Customization includes feature engineering, modeling scheme for partial structure design, fine-tuning by domain data, structural constraints, etc. In sugar and dye materials, "bare" inverse-design without any customization generated extremely diverse structures, totally out of the likelihood for sugar or dye. Those material classes are defined by fragments' types to include, therefore fragment-based structural constrains critically improved the results. Third, to allow a user to customize and tune, a solution should be clear and transparently aligned with an actual user's material knowledge. At the early stage of development, we found that a prototype filled with machine learning terminology (e.g., "latent space," "principal component," etc.) baffled or confused a chemist, decreasing their motivation to use the system. Practical solutions satisfying those requirements have been awaited.

## 7. CONCLUSION

We present a novel AI-driven molecular inverse design service deployed and tested in material companies. Our system is practically useful for materials companies in mainly three reasons; first, feature encoding and structure generation process are algorithm-based, not data-driven, therefore no pre-training with a large dataset is required. Second, the feature space and structure generation process are fully interpretable for human chemists, therefore easy to customize ingesting explicit chemical insights about molecular structures. Third, hierarchical data structures and a clear UI provide a flexible and intuitive workflow. The system is deployed on IBM's and NAGASE's cloud and provided as-as-service to more than 50 chemists. Inverse-design of sugar and dye molecules were carried out by those users, and it was confirmed to outperform human chemists several 10-fold in speed, and the diversity of molecular structures were expanded while still satisfying chemical rationality.

## ACKNOWLEDGMENTS
A part of this work was conducted under IBM Research Frontiers Institute with its member companies; Canon, Hitachi Metals, Honda, JSR, NAGASE, and Samsung (Alphabetical order). We thank them for their collaboration and feedback about our system.